\documentclass[prb,twocolumn,superscriptaddress,showpacs,amsmath,amssymb]{revtex4}

\usepackage{graphicx}
\usepackage{latexsym}
\usepackage{amsmath}
\usepackage{amssymb}
\usepackage{amsfonts}
\usepackage{color}
\usepackage{bm}
\usepackage{verbatim}
\usepackage{color}

\begin{document}
\title{Local density of states around single vortices and vortex pairs:\\
effect of boundaries and hybridization of vortex core states}

\author{A.~S.~Mel'nikov, D.~A.~Ryzhov, M.~A.~Silaev}
\affiliation{Institute for Physics of Microstructures, Russian Academy of
Sciences, 603950 Nizhny Novgorod, GSP-105, Russia}

\date{\today}

\begin{abstract}
The profiles of local density of states (LDOS) around single vortices and
vortex pairs in type-II superconductors are studied taking account of the
interference of quasiparticle waves experiencing Andreev reflection within
the vortex cores, hybridization of vortex core states and normal
reflection at the boundaries or defects. For subgap energy levels these
interference effects reveal themselves in a nontrivial dependence of the
positions of the LDOS peaks on the intervortex distance and sample size:
the peak positions generally do not coincide with the superconducting
phase singularity points. The LDOS profiles are calculated for three
generic examples which can be realized, e.g., in mesoscopic
superconductors: (i) vortex-vortex pair; (ii) vortex positioned near a
flat boundary; (iii) vortex positioned in the center of a superconducting
disk. The resulting evolution of the Andreev interference patterns could
be observable by scanning tunneling spectroscopy techniques in mesoscopic
superconductors or disordered vortex arrays.
\end{abstract}

\pacs{74.25.Jb, 74.25.Qp, 74.78.Na}

\maketitle

\section{Introduction}

Physics of vortex matter in superconductors has been a long standing topic
of intensive research for many years. Over a last decade growing interest
to this topic has been stimulated by technological advances allowing
deeper insight into the structure of vortices as well as investigation of
new exotic vortex states. In particular,  local density of states (LDOS)
measurements with the help of scanning tunneling spectroscopy techniques
have been proven to be an effective tool of experimental study of
electronic structure of Abrikosov vortices\cite{STSHess, STSRoditchev,
STSmore1, STSmore2, RMP-2007}. The observation of the zero-bias anomaly of
tunneling conductance at the center of each vortex in these experiments
clearly confirmed the existence of bound vortex core states predicted by
Caroli, de Gennes and Matricon (CdGM)\cite{CdGM}. The wave functions of
the subgap states are localized inside the vortex core because of the
Andreev reflection of quasiparticles at the core boundary. For each
individual vortex the energy $\varepsilon(\mu)$ of a subgap state varies
from $-\Delta_0$ to $+\Delta_0$ as  one changes the angular momentum $\mu$
defined with respect to the vortex axis. At small energies
$|\varepsilon|\ll\Delta_0$ the spectrum is a linear function of $\mu$:
$\varepsilon(\mu)\simeq-\mu\omega$, where
$\omega\approx\Delta_0/(k_F\xi)$, $\Delta_0$ is the superconducting gap
value far from the vortex axis, $k_F$ is the Fermi momentum, $\xi=\hbar
v_F/\Delta_0$ is the coherence length, $v_F$ is the Fermi velocity, and
$\mu$ is half an odd integer. Note that hereafter we assume the Fermi
surface to be a cylinder and neglect the dependence of the quasiparticle
energy on the momentum component along the vortex axis $z$ considering a
motion of quasiparticles only in $xy$ plane. Introducing a cylindrical
coordinate system $(r,\theta,z)$  and defining an impact parameter as
$b=-\mu/k_F=[{\bf r},{\bf k}_F]\cdot{\bf z}_0/k_F$ the quasiclassical LDOS
inside the isolated vortex core ($r\ll\xi$) can be found as follows:
$N=(k_F/2\pi\xi)\int_0^{2\pi}\delta(\varepsilon-\omega
k_Fr\sin(\theta-\theta_p))d\theta_p$. Here we evaluate the LDOS summing up
over the quasiparticle states at the trajectories characterized by the
orientation angle $\theta_p$. This expression yields a singularity of zero
energy LDOS at the vortex center\cite{Ullah, Maki, IchiokaStar}:
$N=1/(2\pi\omega r\xi)\approx N_0 \xi/r$, where $N_0=(1/2\pi)m/\hbar^2$ is
a normal metal LDOS. Smearing of energy levels due to scattering effects
leads to a reduction of LDOS peak amplitude at the vortex center. However,
the peak in the LDOS distribution survives even in "dirty" limit when a
mean free path is smaller than a coherence length $l<\xi$ (see
Refs.~\onlinecite{golubov, IchiokaDirty}).


The increase in the external magnetic field results in the
decrease in the intervortex distance which should be accompanied
by the overlapping of wave functions describing the quasiparticle
states bound to the neighboring vortex cores. This hybridization
of vortex core states can perturb the CdGM spectra of isolated
vortices and the circular symmetry of the LDOS peaks (see, e.g.,
Ref.~\onlinecite{IchiokaStar}). Note that the LDOS peak anisotropy
may also reflect possible superconducting gap and Fermi surface
anisotropy, and formation of the charge density wave
order\cite{STSHess, STSmore2, IchiokaStar}.

The goal of the present manuscript is to show by the example of mesoscopic
superconductor that the interference of quasiparticle waves trapped in
neighboring vortex cores can result even in more dramatic consequences for
the LDOS profiles around the vortex configurations. These consequences are
not limited just to the distortion of the LDOS peaks at individual vortex
positions caused by the concrete symmetry of the system and/or vortex
configuration. The peak positions themselves appear to be shifted from the
superconducting phase singularity points which are usually considered as
vortex centers. Changing the external magnetic field we can control the
vortex configuration in the sample and switch between different Andreev
interference patterns in LDOS. The distinctive features of the electronic
structure of vortices in mesoscopic samples with sizes of several
coherence lengths are controlled by two key factors: (i) quasiparticle
tunneling between the neighboring vortex cores which becomes extremely
important when the individual vortices merge into the giant (multiquantum)
vortex\cite{Mesovortices} with a winding number larger than unity $M>1$;
(ii) normal scattering of quasiparticles at the sample boundary.

The multiquantum vortex provides a simple illustration of the above
statement regarding the difference in the positions of the LDOS peaks and
phase singularity points. Indeed, the subgap spectrum of such vortex
contains $M$ anomalous spectral branches (per spin
projection)\cite{VolovikBranches}. Each anomalous branch intersects the
Fermi level and contributes to the low-energy DOS. At low energies the
spectrum has the following form\cite{VolovikBranches, multi-spectrum-num,
Melnikov-Vinokur-2002, Janko}:
\begin{equation}
\label{Volovik-spectr}
  \varepsilon_j(\mu)\sim-\frac{\Delta_0}{k_F\xi}(\mu-\mu_{j})\,,
\end{equation}
where the index $j$ enumerates different spectral branches ($1<j<M$),
$-k_F\xi\lesssim\mu_{j}\lesssim k_F\xi$. The LDOS profile corresponding to
the spectrum (\ref{Volovik-spectr}) consists of a set of axially symmetric
ring structures\cite{Melnikov-Vinokur-2002, Janko}. Note that for an even
winding number the anomalous branch crossing the Fermi level at $\mu=0$
(i.e. at zero impact parameter) is absent and, as a result, the LDOS peak
at the vortex center disappears. The splitting of a multiquantum vortex
into the individual vortices should lead to the transformation of LDOS
rings into a set of peaks positioned in the cores of individual vortices.
The initial stage of this LDOS transformation for small intervortex
distances was studied in Ref.~\onlinecite{Melnikov-Vinokur-2002} within a
perturbation approach. A nonperturbative approach which allows to describe
the spectrum transformation accompanying the decay of the multiquantum
vortex has been suggested in Refs.~\cite{MelnikovSilaev, MelnikovPrb2008}.

The second factor which is crucial in mesoscopic superconducting samples
is a quasiparticle scattering at the sample boundaries which comes into
play when vortices approach a superconductor surface or a size of
superconducting sample is small enough. For vortices positioned rather
close to the sample surface the effect of quasiparticle reflection at the
boundary on the spectrum and total DOS was investigated recently in
Refs.~\onlinecite{MelnikovPrl2005, MelnikovPrb2007, MelnikovPrb2008}. The
early stage of transformation of the LDOS profiles for vortices
approaching a flat boundary has been studied numerically in
Ref.~\onlinecite{Schopohl}.

In the present paper we analyze the effect of both the
hybridization of vortex core states and boundary scattering on the
LDOS structures and for this purpose consider three generic
examples: (i) vortex-vortex pair; (ii) vortex positioned near a
flat boundary; (iii) vortex positioned in the center of a
superconducting disk. We will show that the LDOS profiles can be
obtained qualitatively from the spectra found in
Refs.~\onlinecite{MelnikovPrl2005, MelnikovSilaev,
MelnikovPrb2008}. Our qualitative considerations will be confirmed
by the detailed numerical calculations.

This paper is organized as follows. In Sec.~\ref{sec:theory} we give an
overview of the theoretical framework which is employed in this work,
namely the Bogoliubov--de Gennes theory and the quasiclassical Eilenberger
approach.  We discuss the spectrum and LDOS patterns for multi vortex
configurations in Sec.~\ref{subsec:molecules} and address the case of a
vortex in a small mesoscopic cylinder in Sec.~\ref{subsec:cylinder}. We
give our conclusions in Sec.~\ref{sec:summary}.

\section{Model and basic equations}
\label{sec:theory} Our further consideration is based on two approaches:
(i) Bogoliubov--de Gennes (BdG) equations; (ii) quasiclassical Eilenberger
theory. The first one appears to be more transparent for the qualitative
analysis of the interference effects and convenient for numerical
calculation of LDOS profiles around the vortex placed in a superconducting
disk. The Bogoliubov--de Gennes (BdG) equations for particle-- ($u$) and
hole--like ($v$) parts of the wave function have the following form:
\begin{eqnarray}
\label{BdG}
  \nonumber
  -\frac{1}{2m}\left(\hat {\bf p}-\frac{e}{c}{\bf A}\right)^2u+\Delta v
  &=&(\varepsilon+\varepsilon_F) u\,,\\
  \frac{1}{2m}\left(\hat {\bf p}+\frac{e}{c}{\bf A}\right)^2v
  +\Delta^* u&
  =&(\varepsilon-\varepsilon_F) v\,.
\end{eqnarray}
Here $\Delta$ is a gap function, ${\bf A}$ is a vector potential, $\hat
{\bf p}=-i\hbar(\partial/\partial x,\partial/\partial y)$, and ${\bf
r}=(x,y)$ is a radius vector in the plane perpendicular to the vortex
axis. The LDOS can be expressed through the eigenfunctions of the BdG
equation (\ref{BdG}) in the following form (see, e.g.,
Ref.\onlinecite{Shore-Huang-Dorsey-1989}):
\begin{equation}\label{DOS}
  N({\bf r},\varepsilon)=\sum_n
  |u_n({\bf{r}})|^2\delta(\varepsilon-\varepsilon_n)\,,
\end{equation}
where $u_n({\bf{r}})$ is electron component of quasiparticle eigenfunction
corresponding to an energy level $\varepsilon_n$ (we sum over both
positive and negative values $\varepsilon_n$). The eigenfunction has to be
normalized:
  $$
\int\left(|u_n ({\bf r})|^2+|v_n ({\bf r})|^2\right)d^2{\bf r}=1\,.
  $$

In general case BdG equations are rather complicated. A simplification can
be obtained if one considers a quasiclassical approximation, assuming that
the wavelength of quasiparticles is much smaller than the superconducting
coherence length (see, e.g., Ref.~\onlinecite{Bardeen}). Within such an
approximation, quasiparticles move along linear trajectories, i.e.
straight lines along the direction of the quasiparticle momentum ${\bf
k}_F=k_F(\cos\theta_p,\sin\theta_p)$. Generally, the quasiclassical form
of the wave function can be constructed as follows: $(u,v)=e^{i{\bf
k}_F{\bf r}}(\tilde U,\tilde V)$, where $(\tilde U({\bf r}),\tilde V({\bf
r}))$ is a slowly varying envelope function. Then the system (\ref{BdG})
is reduced to the quasiclassical Andreev equations:
\begin{align}
\label{quasiclass}
  \frac{\hbar\mathbf{k}_F}{m}\left(-i\hbar\nabla+
  \frac{e}{c}\mathbf{A}\right)\tilde U+\Delta\tilde V
  =\varepsilon\tilde U\,,
  \notag\\
  \frac{\hbar\mathbf{k}_F}{m}\left(i\hbar\nabla+
  \frac{e}{c}\mathbf{A}\right)\tilde V+\Delta^*\tilde U
  =\varepsilon\tilde V\,,
\end{align}
which are defined at the linear trajectories determined by the direction
of the quasiparticle momentum ${\bf k}_F$.

The quasiclassical approximation allows to develop a powerful method for
calculation of the LDOS based on the solution of Eilenberger equations for
quasiclassical propagator along the trajectories\cite{Eilenberger}. For
numerical treatment of these equations we follow the
Refs.~\onlinecite{Maki, schopohl_cm} and introduce a Ricatti
parametrization for the Green function. The essence of this method is a
mathematical trick which allows to solve two first order Ricatti equations
instead of fourth-order system of Eilenberger equations. Following
Refs.~\onlinecite{Maki, Schopohl, schopohl_cm} one can obtain:
\begin{align}
\label{eq:eilenberger}
  \hbar v_F\frac{\partial}{\partial x'}a(x')
  &+[2\tilde{\omega}_n+\Delta^*a(x')]a(x') - \Delta = 0 ,\notag\\
  \hbar v_F\frac{\partial}{\partial x'}b(x') &-
  [2\tilde{\omega}_n + \Delta b(x')]b(x') + \Delta^* = 0,
\end{align}
where $x'=({\bf k}_F{\bf r})/k_F=r\cos(\theta_p-\theta)$ is the coordinate
along trajectory, $\Delta({\bf r})=|\Delta|e^{i\Phi}$, $\Phi({\bf r})$ is
a superconducting phase, $i\tilde{\omega}_n=i\omega_n+m{\bf v}_F{\bf v}_s$
is a Doppler-shifted Matsubara frequency, $\omega_n=\pi T(2n+1)$ and
  $$
{\bf v}_s=\frac{1}{2m}\left(\hbar\nabla\Phi-\frac{2e}{c}{\bf A}\right)
  $$
is a gauge-invariant superfluid velocity. The LDOS may be expressed
through the scalar coherence functions $a$ and $b$ as
follows\cite{Schopohl}
\begin{align}
\label{eq:LDOS}
  N({\bf r},\varepsilon) = \int^{2\pi}_0 \frac{d\theta}{2\pi}
  \text{Re}\left\{\frac{1-ab}{1+ab}\right\}_{i\omega_n \to \varepsilon+i\nu},
\end{align}
where $\varepsilon$ is the quasiparticle energy measured from Fermi level
and $\nu$ is a parameter which accounts for an elastic scattering.
Throughout this paper we will assume the simplest model\cite{clem} for the
gap function distribution around an isolated vortex positioned at the
origin $\Delta({\bf r})=\Delta_0f_1({\bf r})$, where
\begin{equation}
\label{Clem-profile}
  f_1({\bf r})=\frac{x+iy}{\sqrt{x^2+y^2+\xi_v^2}}
  =\frac{r e^{i\theta}}{\sqrt{r^2+\xi_v^2}}
\end{equation}
with the core size $\xi_v=\xi$. To describe the system of two
singly--quantized vortices positioned at ${\bf r}=\pm{\bf a}/2=\pm(a/2,0)$
we fit the gap function as follows:
\begin{eqnarray}
\label{2VortMolecGap}
  \nonumber
  \Delta({\bf r})=\Delta_0f_1({\bf r}-{\bf a}/2)f_1({\bf r}+{\bf a}/2)\,.
\end{eqnarray}

The image method allows us to reduce the problem of a vortex positioned at
the point ${\bf r}=(a/2,0)$ near a flat boundary $x=0$ to the one
describing a vortex-antivortex pair with the antivortex situated at ${\bf
r}=-(a/2,0)$. For the latter case we choose the gap function in the form:
\begin{eqnarray}
\label{VaVVortMolecGap}
 \nonumber
 \Delta({\bf r})=\Delta_0f_1({\bf r}-{\bf a}/2)f_1^*({\bf r}+{\bf a}/2)\,.
\end{eqnarray}

Certainly, the above simplest approximation for the gap function
might not be enough to describe closely spaced vortex--vortex or
vortex--antivortex pairs. However, the resulting LDOS patterns
obtained below  are mainly controlled only by the distance between
the phase singularity points  and appear to be weakly sensitive to
the concrete profiles of the order parameter absolute value. The
key features of the LDOS patterns appear to hold even in the limit
of the zero $\xi_v$ value, i.e., even for phase vortices without
any suppression of the gap value inside the cores.

Considering hereafter only the vortex configurations of the finite size
comparable with several coherence lengths we neglect the vector potential
which is known to give only a moderate renormalization of the interlevel
spacing $\omega$ (see Ref.~\onlinecite{BrunHansen}).

\section{Spectrum and local density of states}
\label{sec:results}

\subsection{Vortex molecules}
\label{subsec:molecules}

We now proceed with the calculation of LDOS profiles for two systems: (i)
vortex--vortex pair and (ii) vortex positioned near the flat
superconducting boundary. The image method discussed in
Ref.~\onlinecite{MelnikovPrb2008} allows us to show that the latter case
is equivalent to the vortex-antivortex pair in the bulk provided we choose
the distance between the vortex and antivortex to be two times larger than
the distance from the vortex to the flat surface in the original problem.
To elucidate our main results we start from a qualitative description of
the quasiparticle spectra following Ref.~\onlinecite{MelnikovPrb2008}.

Let us consider two vortices oriented along the $z$ axis. In the plane
$(xy)$ the corresponding phase singularity points (or vortex centers) are
positioned at  ${\bf r}_{1,2}=(\pm a/2,0)$. For quasiparticles propagating
along the classical trajectories parallel to ${\bf k}_F$ we introduce the
angular momenta $\mu=[{\bf r},{\bf k}_F]\cdot{\bf z_0}=k_F
r\sin(\theta_p-\theta)$ and $\tilde\mu_i=\mu-[{\bf r}_i,{\bf
k}_F]\cdot{\bf z}_0$ defined with respect to the $z$--axis passing through
the origin and with respect to the $i$-th vortex axis passing through the
point ${\bf r}_i$, correspondingly. Neglecting the quasiparticle tunneling
between the vortex cores the wave function  can be found as a
superposition of two states localized at different vortices and having
close energies: $\varepsilon_{v1}=-\omega[\mu-(k_F a/2)\sin\theta_p]$ and
$\varepsilon_{v2}=-\omega[\mu+(k_F a/2)\sin\theta_p]$. The transformation
of the quasiclassical spectrum occurs due to the overlapping of the
corresponding wave functions and can be analysed using a standard quantum
mechanical approach  describing a two--level system\cite{LL-III}, which
yields the secular equation
\begin{equation}
\label{secularVV}
  (\varepsilon-\varepsilon_{v1})(\varepsilon-\varepsilon_{v2})
  =(\delta\varepsilon)^2\,,
\end{equation}
and the resulting splitting of isoenergetic lines near the degeneracy
point in the $\mu-\theta$ plane ($\theta_p=0$ for our case)
\begin{equation}
\label{orbit1}
  \omega\mu=-\varepsilon\pm\sqrt{\omega^2(k_F a/2)^2\theta_p^2
  +(\delta\varepsilon)^2}\,.
\end{equation}
The tunneling of quasiparticles between the vortex cores is determined by
the exponentially small overlapping of wave functions localized near the
cores and results in the splitting of the energy levels:
$\delta\varepsilon\sim\Delta_0\exp(-a/\xi)$. Thus, the estimate for the
splitting $\delta\mu\simeq\delta\varepsilon/\omega$ of isoenergetic lines
in the $(\mu,\theta_p)$--plane reads (see Eq. (\ref{orbit1})):
\begin{equation}
\label{delta-mu}
  \delta \mu(a)\sim k_F\xi\exp\left(-\frac{a}{\xi}\right)\,.
\end{equation}

In the case of a vortex-antivortex pair the non-interacting states
localized at different vortices have the energies
$\varepsilon_v=-\omega[\mu+(k_F a/2)\sin\theta_p]$ and
$\varepsilon_{av}=\omega[\mu-(k_F a/2)\sin\theta_p]$. Taking into account
the overlapping of the corresponding wave functions states we obtain the
secular equation (\ref{secularVV}) with $\varepsilon_{v1}=\varepsilon_v$
and $\varepsilon_{v2}=\varepsilon_{av}$. Therefore, the quasiclassical
orbits near the degeneracy point $\theta_p=0$ are defined by the following
equation:
\begin{equation}
\label{orbit2}
  \varepsilon=\pm\sqrt{(\omega\mu)^2+(\delta \varepsilon)^2}
  -\omega(k_F a/2)\theta_p\,.
\end{equation}
The classically forbidden angular domain at $\varepsilon=0$ has the width
$\delta\theta_p=4\delta\varepsilon/(\omega k_F a)$. One can assume that
the appearance of such classically forbidden domain explains the deep
structure in the local DOS profile observed numerically in
Ref.~\onlinecite{Schopohl} for a vortex near the flat boundary of an
$s$--wave superconductor. The classically forbidden angular domain results
in the suppression of the overall DOS\cite{MelnikovPrb2008} and we show
below that this mechanism is responsible for the anomalous spectrum branch
disappearance when the vortex exits the sample.

\begin{figure}[htb]
\centerline{\includegraphics[width=1.0\linewidth]{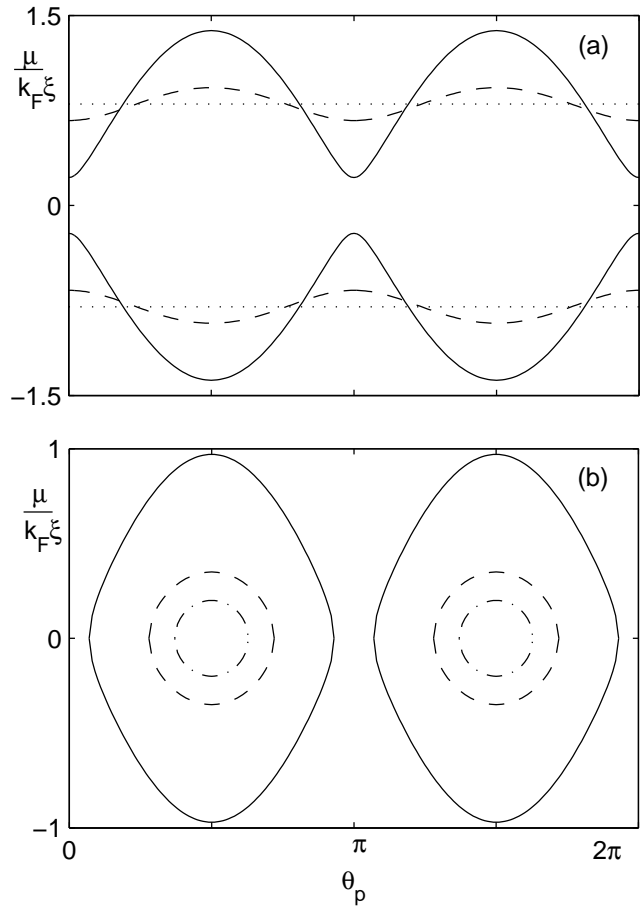}}
\caption{\label{Orbit} Quasiclassical orbits in the $\mu-\theta_p$ plane
corresponding to zero energy level. Upper panel (a): two vortices at
distances $a=2.5\xi$ (solid line) and $a=1.5\xi$ (dashed line). Dotted
lines correspond to the case of a doubly quantized vortex, i.e. to $a=0$. Lower panel
(b): vortex-antivortex system with $a=2.5\xi$ (solid line), $a=1.7\xi$
(dashed line) and $a=1.6\xi$ (dashdot line).}
\end{figure}

To study the transformation of spectrum in the entire range of intervortex
distances we solve numerically the quasiclassical Andreev equations
(\ref{quasiclass}) to find isoenergetic lines in $\mu -\theta_p$ plane.
The resulting quasiclassical orbits for a vortex--vortex pair with the
intervortex distances $a=2.5\xi$ and $a=1.5\xi$ are shown in
Fig.~\ref{Orbit}a.
The dotted lines show the orbits for $a=0$, i.e., for a doubly quantized
vortex. The splitting of the quasiclassical orbits grows with the decrease
in the size of the vortex molecule both for the two-vortex and
vortex-antivortex configurations. When the intervortex distance is of the
order $\xi$ the energy splitting becomes so large
($\delta\varepsilon\sim\Delta_0$) that the simplified equations
(\ref{orbit1}),(\ref{orbit2}) are no longer valid and the spectrum is
disturbed in the whole angular domain $0<\theta_p<2\pi$. One can clearly
observe that with the decrease in the intervortex distance the CdGM orbits
$b=\pm (a/2)\sin\theta_p$ corresponding to zero energy in isolated
vortices transform into two lines $\mu=\pm\mu_0$ which correspond to the
spectrum of a doubly quantized vortex. Comparing the isoenergetic lines
with the unperturbed CdGM ones one can see that the zero energy
quasiclassical trajectories do not pass through the vortex center. It
means that the LDOS peaks are necessary shifted from  the vortex centers
as the intervortex distance becomes small enough. This phenomenon can be
understood considering the following qualitative arguments. Indeed, when
the intervortex distance is large $a\gg\xi$ the equilibrium size of the
vortex pair is determined by the stability condition yielding the zero
value of Lorentz force acting on each vortex: $ {\bf F}_L\propto\phi_0
[{\bf z}_0,{\bf v}_s]=0$, where $\phi_0$ is a magnetic flux piercing the
vortex and ${\bf v}_s$ is a a superfluid velocity induced in a given
vortex core by other vortices and screening currents flowing along the
boundaries. Thus, the local Doppler shift of the quasiparticle energy
levels $\varepsilon_d=\hbar{\bf k}_F{\bf v}_s$ appears to vanish for
trajectories passing through the vortex center. As a result, the position
of the LDOS peaks coincide with the phase singularity points which are
usually defined as vortex centers. However when the intervortex distance
is comparable with (or less than) $\xi$ the part of the superfluid
velocity  induced by the neighboring vortex becomes strongly inhomogeneous
inside the core region and the ${\bf v}_s$ value at a given vortex center
diverges as $v_s\sim1/a$. As a consequence, the equilibrium size of the
pair is no more determined by the condition of zero ${\bf v}_s$ at the
vortex centers. The resulting nonvanishing Doppler shift for the
trajectories passing through the vortex centers suppresses the LDOS at
these points. Thus, for small $a$ the LDOS peak positions do not coincide
with the vortex centers.

The Fig.~\ref{Orbit}b corresponds to the vortex-antivortex pair with the
intervortex distances $a=2.5\xi$, $a=1.7\xi$, and $a=1.6\xi$. One can see
that contrary to the case of the vortex-vortex pair the decrease in the
intervortex distance in this system leads to a rapid shrinking of the
isoenergetic orbits. When the distance $a$ is small enough all the
trajectories corresponding to the zero energy  are characterized by the
$\mu$ and $\theta_p$ values  close to $\mu=0$ and $\theta_p=\pi (n+1/2)$,
where $n$ is integer. For the intervortex distance $a/\xi=1.5$ the
quasiclassical orbits  completely shrink to these points in $\mu-\theta_p$
plane. In the real space the corresponding trajectories pass through the
point in the middle between the vortex and antivortex and are
perpendicular to the line connecting the phase singularity points. Thus,
the two peaks of LDOS should shift from the vortex centers to the middle
point and merge into one peak as the vortex and antivortex approach each
other.

\begin{figure}[hbt]
\centerline{\includegraphics[width=1.0\linewidth]{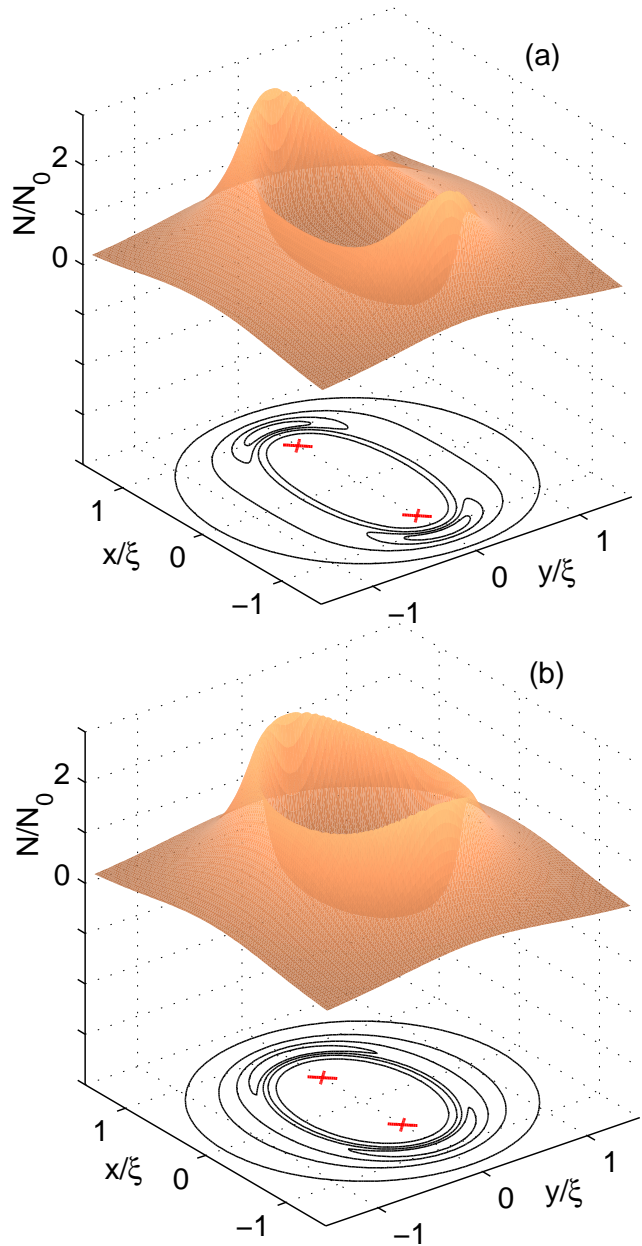}}
\caption{\label{LDOSvv} (Color online) The normalized LDOS profile
in a two--vortex system for $a/\xi=1.5$ (a) and $a/\xi=1$ (b)
calculated for $\nu=0.06\Delta_0$. Positions of vortices are
marked by red crosses.}
\end{figure}

\begin{figure}[hbt]
\centerline{\includegraphics[width=1.0\linewidth]{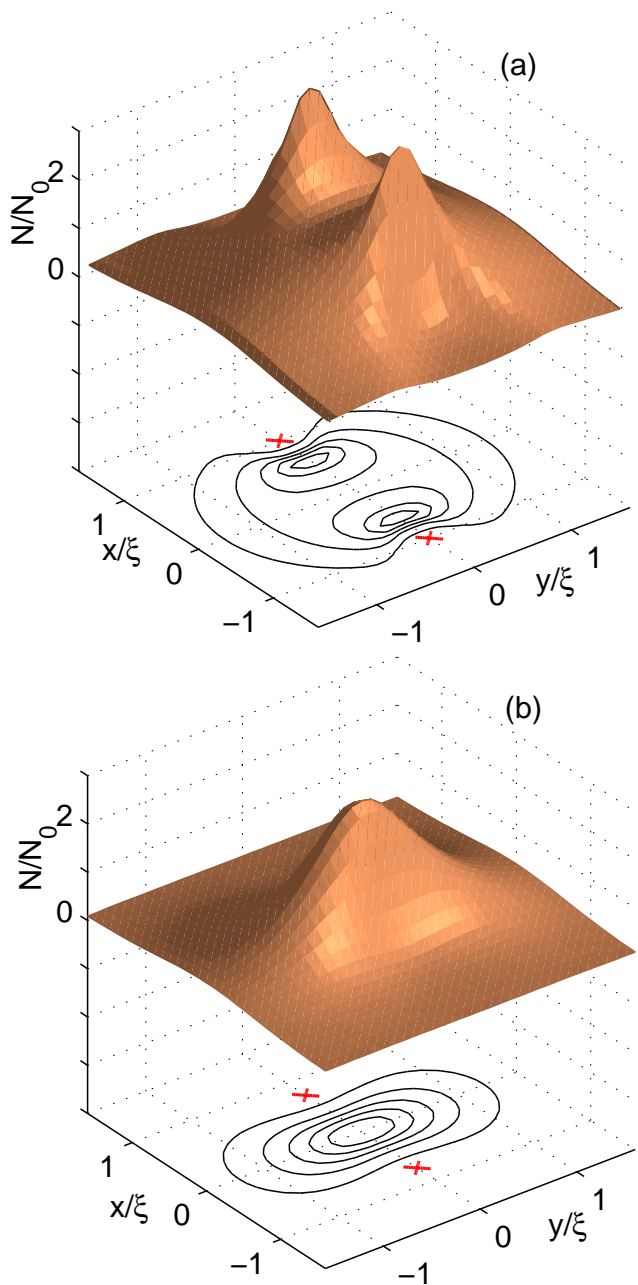}}
\caption{\label{LDOSvav} (Color online) The normalized LDOS
profile in a vortex--antivortex system for $a/\xi=2$ (a) and
$a/\xi=1.5$ (b) calculated for $\nu=0.06\Delta_0$. Positions of
vortices are marked by red crosses.}
\end{figure}

For a detailed study of these effects we calculate the LDOS profiles for
two--vortex and vortex--antivortex configurations by solving numerically
the Eilengerger equations (\ref{eq:eilenberger}) and applying the
expression for the LDOS (\ref{eq:LDOS}). The results are shown in
Fig.~\ref{LDOSvv} for the vortex--vortex pair with $a=1.5\xi$ (upper
panel) and $a=\xi$ (lower panel) and in Fig.~\ref{LDOSvav} for a
vortex--antivortex pair with $a/\xi=2$ (upper panel) and $a=1.5\xi$ (lower
panel).
In Fig.~\ref{LDOSvv} it is shown how the LDOS profile gradually transforms
from a two-peaked structure corresponding to two isolated vortices (upper
panel) to the axially symmetric ring for a doubly quantized vortex. The
positions of the phase singularity points are shown on the contour plot by
red crosses. It can be seen that even when the distance between the
vortices exceeds the size of the vortex core the positions of the LDOS
peaks do not coincide with the vortex centers. The distance between the
LDOS peaks reduces slower than the size $a$ of the vortex molecule and
finally the peaks are smeared into the anisotropic ring structure (see
Fig.~\ref{LDOSvv}b). The anisotropy vanishes and the ring becomes axially
symmetric when vortices merge into a doubly quantized vortex.

The evolution of the LDOS profile in a vortex-antivortex pair follows a
different scenario (see Fig.~\ref{LDOSvav}). When the distance between
vortices is large enough there appear deep structures at the LDOS peaks
(see Fig.~\ref{LDOSvav}a). The peaks themselves are slightly shifted
towards the middle point at the line connecting the vortex and antivortex
centers. Below the critical intervortex distance $a_c/\xi\sim 1.7$ the two
peaks merge into one peak at the middle point (see  Fig.~\ref{LDOSvav}b).
As we discussed above this phenomenon is explained by the shrinkage of the
quasiclassical orbits in the $\mu - \theta_p$ plane. When the vortex and
antivortex come closer the amplitude of the LDOS peak is reduced and
finally it disappears completely when the vortex and antivortex merge and
annihilate. Mapping the solution of the vortex--antivortex problem on the
one for a vortex near the flat surface we conclude that the zero bias LDOS
peak appears to be positioned exactly at the surface provided the distance
from the vortex to the surface becomes less than $0.85\xi$.

\subsection{Vortex in mesoscopic cylinder}
\label{subsec:cylinder}

In this subsection we study another generic example illustrating the
effect of the normal quasiparticle reflection at the sample boundary on
the LDOS profiles in the vortex state. We consider a singly quantized
vortex situated at the center of a superconducting disk of rather small
radius $R$ comparable with the coherence length. In this case a
constructive interference of quasiparticle waves reflected from the
boundary is known to result in mesoscopic oscillations of energy
levels\cite{MelnikovPrl2005}:
\begin{equation}
\label{SpectrumDisc}
  \varepsilon(\mu)=-\omega\mu -\delta\sin(2k_FR-\pi \mu)\,,
\end{equation}
where $\delta\sim \Delta_0\exp(-2R/\xi)$ and
$|\varepsilon(\mu)|\ll\Delta_0$. The remarkable fact is that at low
energies these levels can be combined into two groups corresponding to odd
and even values of the integer index $\mu+1/2$. A spacing between the
energy levels belonging to one group is $2\omega$ while the energy shift
between the different groups is $2\delta|\cos(2k_F R)|$. Analogous shift
of vortex core levels defined by the factor $\exp(-R/\xi)$ has been
observed recently for a vortex-antivortex pair on a sphere with a finite
radius $R$ in $p_x+ip_y$ superconductors\cite{Fertig}. Note that for a
small superconducting disc with $R<R_c\sim (\xi/2)\ln (k_F\xi)$ we have
$\delta\gg\omega$. Therefore the two groups of energy levels can be
considered as continuous branches if the energy discreteness with the
scale $\omega$ is neglected. In this approximation one can use a
quasiclassical expression $\mu=-k_Fr\sin(\theta-\theta_p)$ and obtain the
low energy branches in the form:
\begin{equation}
\label{SpectrumDisc-q}
  \varepsilon_{1,2}
  =\omega k_Fr\sin(\theta-\theta_p)\pm\delta\cos(2k_F R)\,.
\end{equation}
Thus, for low energies $\varepsilon\ll\Delta_0$
the quasiclassical expression for the LDOS near the vortex center reads:
\begin{equation}
\label{LDOSDisc0}
  N(r,\varepsilon)=\frac{k_F}{2\pi\xi}\sum_{j=1,2}\int_0^{2\pi}\delta
  (\varepsilon-\varepsilon_{j})d\theta_p\,.
\end{equation}
The Eq.(\ref{LDOSDisc0}) can be derived from the general expression for
LDOS (\ref{DOS}) by setting the amplitude of the wave function $|\tilde
U_n|^2={\rm const}$ which is a good approximation at the small distances
$r\ll\xi$ from the vortex center. Evaluating this expression for the LDOS
we obtain:
\begin{eqnarray}
\label{LDOSDisc}
  N(r,\varepsilon=0)=N_0\frac{\xi}{\sqrt{r^2-r_0^2}}\,,
  \quad {\rm for}\ \ r>r_0\,,
  \nonumber\\
  N(r,\varepsilon=0)=0\,, \quad {\rm for}\ \  r<r_0\,,
\end{eqnarray}
where
\begin{equation}
  r_0=\frac{\delta|\cos(2k_F R)|}{\omega k_F}
  \sim\xi|\cos(2k_F R)|e^{-2R/\xi}\,.
\end{equation}
Thus, the zero-bias peak of the LDOS at the vortex center which exists in
an isolated vortex transforms into a ring structure which is similar to
the LDOS pattern for a doubly quantized vortex. The only difference is
that the radius of this ring is exponentially small  comparing to the one
in a doubly quantized vortex: $r_0/\xi\propto\exp(-R/\xi)\ll1$. Therefore,
this splitting of the LDOS peak in tunneling spectroscopy experiments
should be strongly affected by different smearing effects like disorder
scattering or finite temperature. In particular, for the observation of
the above effect we should assume the elastic mean free path to exceed the
characteristic size of the LDOS ring: $l\geq\xi\exp(-R/\xi)$. Note that
this condition is not very restrictive because it can be fulfilled even in
"dirty" superconductors with $l\ll\xi$ because typically
$\exp(-R/\xi)\ll1$. The effect of smearing due to the finite temperature
is controlled by the parameter $T/\delta$.

\begin{figure}[hbt]
\centerline{\includegraphics[width=1.0\linewidth]{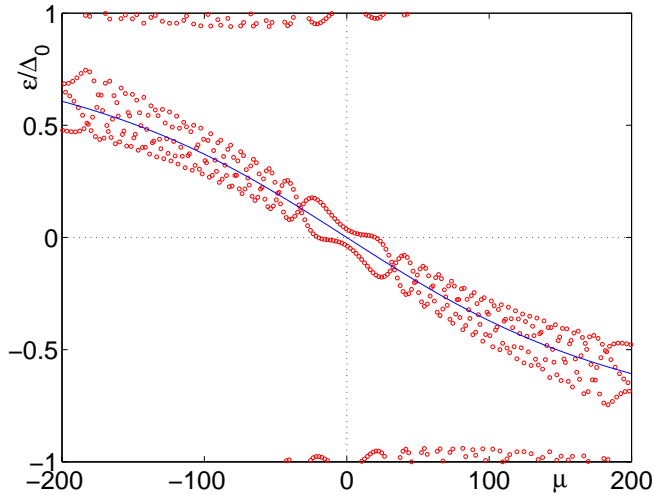}}
 \caption{\label{FigSpectrumDisc}
(Color online) Quasiparticle spectrum vs $\mu$ for a singly quantized
vortex in a disk of the radius $R=2\xi$. The CdGM spectrum is shown by the
solid line. We choose here $k_F\xi=200$.}
\end{figure}

To take into account the finite temperature effects and investigate the
LDOS profiles in the whole range of energies and distances from the vortex
core we solve numerically the BdG system (\ref{BdG}) for a vortex in
mesoscopic cylinder. We use the same method which was successfully applied
to study the mesoscopic oscillations of vortex core
levels\cite{MelnikovPrl2005}, the spectrum of multiquanta vortices in
mesoscopic superconductors\cite{MelnikovPrb2008} and the heat transport
along vortex lines\cite{MelnikovPrb2007}. The resulting spectrum as a
function of the angular momentum is shown in Fig.~\ref{FigSpectrumDisc}
which clearly demonstrates the splitting of the anomalous CdGM branch at
low energies caused by the boundary effects.
For large energies $\varepsilon\sim\Delta_0$ the spectrum is not well
described by the Eq.(\ref{SpectrumDisc}).

\begin{figure}[!hbt]
\centerline{\includegraphics[width=1.1\linewidth]{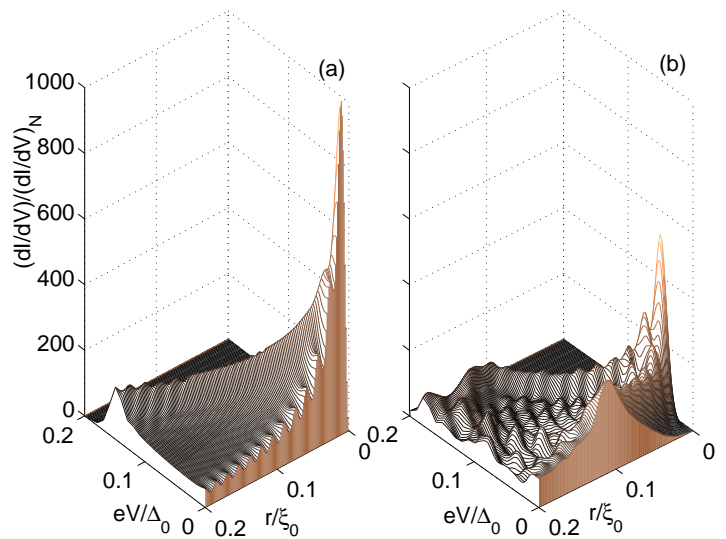}}
 \caption{\label{dIdV_M=1}
(Color online) Three-dimensional plot of the normalized local differential
conductance $dI/dV$ as a function of voltage ($eV$) and distance from the
vortex center ($r$) in disks with a large radius $R=7\xi$ (a) and small
radius $R=2\xi$ (b) for $T/\Delta_0=0.005$. We choose here $k_F\xi=200$.}
\end{figure}

Using the spectrum and the expression (\ref{DOS}) for the LDOS we can find
the local differential conductance as a function of voltage $V$:
  $$
  \frac{dI}{dV}=\left(\frac{dI}{dV}\right)_{N}
  \int_{-\infty}^{+\infty}\frac{N({\bf r},\varepsilon)}{N_0}
  \frac{\partial f(\varepsilon-eV)}{\partial V}\,d\varepsilon\,,
  $$
where $(dI/dV)_{N}$ is a conductance of the normal metal junction and
$f(E)=[1+\exp(E/T)]^{-1}$ is a Fermi distribution function. The typical
plots of the differential conductance are shown in Fig.~\ref{dIdV_M=1} for
different disk radii.
The zero bias peak at the vortex center is clearly seen for a large disk
radius when the normal scattering at the surface can be neglected. The
decrease in the disk radius results in the shift of this conductance peak
to higher voltages and formation of ring structure of local zero-bias
conductance with finite radius $r\sim r_0$.

Note, that normal scattering of quasiparticles at the boundaries can be
also important for multiquanta vortex configurations, e.g., vortex pairs,
trapped in rather small samples. In our consideration of the LDOS profiles
for a vortex-vortex pair in the previous subsection we have neglected the
normal scattering at the sample surface assuming the vortex centers to be
situated rather far from the boundaries. However, from our consideration
of the vortex -- antivortex problem one can expect that the effect of
boundary scattering on the LDOS becomes important only when the distance
from the vortex center to the sample edge is comparable with $\xi$. In
this case the LDOS profiles in a vortex pair are affected by both the
hybridization of the CdGM states and normal scattering. Considering the
particular case of a doubly quantized vortex one can expect the boundary
scattering to result in the spectrum oscillations (see Fig.~6) similar to
the ones in a singly quantized vortex.
\begin{figure}[htb]
\centerline{\includegraphics[width=1.0\linewidth]{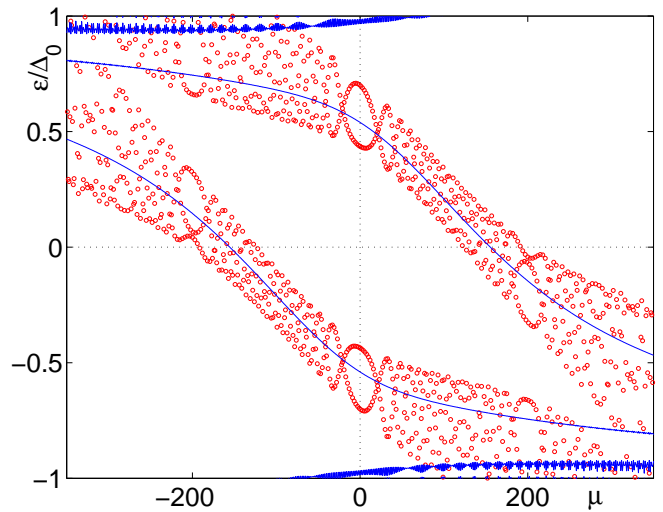}}
 \caption{\label{FigSpectrumDisc2}
(Color online) Quasiparticle spectrum vs $\mu$ for a doubly quantized
vortex in a disk of the small radius $R=2\xi$. The spectrum for a doubly
quantized vortex in a disk of the large radius $R=7\xi$ is shown by the
solid line. We choose here $k_F\xi=200$.}
\end{figure}
These oscillations should cause the broadening of the ring in the LDOS
pattern with the typical broadening scale $r_0$. The same effect should
result in the broadening of the anisotropic ring structure for a vortex
pair of a finite size.

\section{Conclusion}
\label{sec:summary}

To summarize, we have investigated the distinctive features of the LDOS
profiles specific both for the vortex states in mesoscopic superconducting
samples and for vortex arrays in bulk superconductors. These profiles are
shown to be strongly affected by the Andreev interference effects and the
normal scattering at the sample boundaries. In particular, we find that in
contrast to the case of isolated vortices far from the boundaries the
positions of peaks in the LDOS profiles in vortex configurations in small
size samples are shifted from the phase singularity points. Taking a
generic example of the two vortex system, we have considered the evolution
of the LDOS profile which accompanies the merging of two vortices and
appearance of the doubly quantized vortex. In this case the distance
between LDOS peaks reduces slower than the intervortex distance $a$.
Finally, when $a$ is smaller than the vortex core size the two-peaked
structure of LDOS transforms into an anisotropic ring. The anisotropy
vanishes when the vortex positions coincide and a doubly quantized vortex
is formed.

Earlier (see Ref.~\onlinecite{MelnikovPrb2008}) it was shown  that the
spectrum of a single vortex placed near the flat surface is transformed
analogously to the vortex-antivortex system and when the distance is of
the order of the vortex core size, the interlevel spacing in the vortex
spectrum becomes larger than the CdGM value. This effect was argued to
lead to the disappearance of the anomalous spectrum branch when the vortex
approaches the surface. In the present paper we have confirmed this
prediction by calculating  LDOS profile for the entire range of distances
from the vortex to surface. We have found an amazing effect for such
system: as the vortex approaches the surface the zero-bias peak of LDOS
shifts from the vortex center to the surface point positioned at the
minimal distance to the vortex center. For the particular vortex core
model that we have used in this paper the shift of the LDOS peak occurs
when the distance from vortex to surface is $0.85\xi$. Experimentally the
LDOS peaks positioned very close to the superconductor boundary have been
recently observed in vortex state of tungsten (W) based thin films by
scanning tunneling microscopy/spectroscopy techniques\cite{Suderow-2009}.
Note that in the latter experiment LDOS vortex peaks have been studied
near the superconductor/normal metal (Au) interface. We suppose that the
normal scattering of quasiparticles at such boundary caused by either the
Fermi velocity mismatch or some surface barrier can result in the
transformation of the LDOS profiles analogous to the one discussed above
for a superconductor/insulator boundary.

Also we have investigated a measurable consequence of the giant mesoscopic
oscillations of vortex core levels in finite size superconducting
systems\cite{MelnikovPrl2005}. Considering the simplest case of a vortex
positioned at the center of a superconducting cylinder with a radius $R$
we have shown that the zero-bias peak of the local differential
conductance at the vortex center transforms into a ring structure of the
radius of the order of $r_0\sim\xi e^{-2R/\xi}$. We expect that the
unusual behavior of the LDOS profiles which we have investigated could be
observable in scanning tunneling microscopy/spectroscopy experiments in
mesoscopic superconductors which are now in the focus of interest (see,
e.g., Refs.~\onlinecite{Suderow-2009, Zalalutdinov-Fujioka-2000,
Roditchev-2008}).

Certainly, the Andreev interference patterns discussed above should be
smeared by the disorder effects which are controlled by $\nu$ parameter
within the Eilenberger theory. The resulting smearing of the LDOS profiles
can be extremely important, e.g., for amorphous superconducting samples
with rather small mean free path. Still we expect that the shift of the
LDOS peaks from vortex singularity points caused by the Doppler effect
will be observable even in this limit of strong disorder.

\section{Acknowledgements}
It is our pleasure to thank H.\ Suderow and N.\ B.\ Kopnin for stimulating
discussions, and D.\ Roditchev for correspondence. This work was
supported, in part, by Russian Foundation for Basic Research, by the
Program ``Quantum Physics of Condensed Matter'' of RAS, and by the
``Dynasty'' Foundation.

\end{document}